\documentclass[aps,prb,floatfix,twocolumn]{revtex4}
\usepackage{amsmath}
\usepackage{bm}
\usepackage{euscript}
\usepackage{graphicx}
\graphicspath{{./Figures/}}

\def\deg{\mbox{$^{\circ}$}}

\def\ea0{\mbox{ $ea_0$}}

\def\mF0{\mbox{ $\mu\Phi_0$}}
\def\mF0rtHz{\mbox{ $\mu\Phi_0/\sqrt{\rm Hz}$}}

\begin{document}
\title{Observation of the thermal Casimir force}
\author{A. O. Sushkov}
\email{alex.sushkov@yale.edu}
\affiliation{Yale University, Department of Physics, P.O. Box 208120, New Haven CT 06520-8120, USA}
\author{W. J. Kim}
\affiliation{Dept. of Physics, Seattle University, 901 12th Avenue, Seattle, WA 98122, USA}
\author{D. A. R. Dalvit}
\affiliation{Theoretical Division MS B213, Los Alamos National Laboratory, Los Alamos, NM 87545, USA}
\author{S. K. Lamoreaux}
\affiliation{Yale University, Department of Physics, P.O. Box 208120, New Haven, CT 06520-8120, USA}

%
%

\date{\today}

\maketitle

\textbf{Quantum theory predicts the existence of the Casimir force between macroscopic bodies, due to the zero-point energy of electromagnetic field modes around them. This quantum fluctuation-induced force has been experimentally observed for metallic and semiconducting bodies, although the measurements to date have been unable to clearly settle the question of the correct low-frequency form of the dielectric constant dispersion (the Drude model or the plasma model) to be used for calculating the Casimir forces.  At finite temperature a thermal Casimir force, due to thermal, rather than quantum, fluctuations of the electromagnetic field, has been theoretically predicted long ago. Here we report the experimental observation of the thermal Casimir force between two gold plates. We measured the attractive force between a flat and a spherical plate for separations between 0.7~$\mu$m and 7~$\mu$m. An electrostatic force caused by potential patches on the plates' surfaces is included in the analysis. The experimental results are in excellent agreement (reduced $\chi^2$ of 1.04) with the Casimir force calculated using the Drude model, including the $T=300$~K thermal force, which dominates over the quantum fluctuation-induced force at separations greater than 3~$\mu$m. The plasma model result is excluded in the measured separation range.}

There are four known fundamental forces: electromagnetism, gravity, weak, and strong interactions. The weak and strong interactions manifest themselves on length scales on the order of the size of a nucleus, at larger distances electromagnetism and gravity prevail. It may therefore come as a surprise that two macroscopic non-magnetic bodies with no net electric charge (or charge moments) can experience an attractive force much stronger than gravity. This force was predicted by Hendrik Casimir in the late 1940s, and now bears his name~\cite{Casimir1948}. The existence of this force is one of the few direct macroscopic manifestations of quantum mechanics, others are superfluidity, superconductivity, kaon oscillations, and the black body radiation spectrum.


The first experimental test of Casimir's prediction came within a few years~\cite{Sparnaay1958}, but the first ``precision'' measurements were made in the late 1990s~\cite{Lamoreaux1997}, leading to renewed interest in this subject, both for experiment and theory~\cite{Lamoreaux2007,Milonni1994}.
Casimir's result, which he obtained by adding up the zero-point energies of the electromagnetic field modes of the cavity comprising two perfectly-conducting flat plates of area $A$ separated by distance $d$, gives the force between the two plates at zero temperature:
\begin{align}
\label{eq:CasFlat}
F^{(T=0)}_{\rm C(||)} = \frac{\pi^2\hbar c}{240}\frac{A}{d^4}.
\end{align}
Note the presence of the Planck's constant $\hbar$, which indicates that this is a quantum effect, which vanishes in the classical limit. With some notable exceptions~\cite{Bressi2002}, most Casimir force experiments, including ours, are performed with one flat plate and one spherical plate, to avoid the difficulty of aligning two flat plates to be parallel to within fractions of a microradian. In the sphere-plane configuration, the zero-temperature Casimir force for perfect conductors can be deduced from equation~(\ref{eq:CasFlat}) using the proximity force approximation (valid when $d\ll R$)~\cite{Derjaguin1934,Blocki1977}:
\begin{align}
\label{eq:CasSpherePlane}
F^{(T=0)}_{\rm C} = \frac{2\pi^3\hbar c}{720}\frac{R}{d^3},
\end{align}
where $R$ is the radius of curvature of the spherical plate, and $d$ is defined as the distance between the flat plate and the closest point on the spherical plate.

Real-world plates, however, are never perfectly conducting, but instead are characterized by a complex permittivity $\epsilon(\omega) = \epsilon'(\omega) + i\epsilon''(\omega)$, a material-dependent function of frequency $\omega$. This modifies the zero-temperature Casimir force, which now has to be calculated in the framework of the Lifshitz theory~\cite{Lifshitz1956}, based on the computation of the electromagnetic field stress tensor, taking into account the correlated fluctuating charges and currents in the plates. The Lifshitz theory has been tested in experiments with $<200$~$\AA$-thick liquid-helium films on cleaved crystal surfaces~\cite{Sabisky1973}, and measurements of the short-range forces between plates of a number of metallic~\cite{Mohideen1998,Chan2001,Decca2003} and semiconducting~\cite{Kim2009} materials have been found to be in reasonable agreement with the Lifshitz theory at plate separations less than 1~$\mu$m. Repulsive forces have been observed between metallic plates immersed in fluids, also in agreement with the Lifshitz theory~\cite{Feiler2008,Munday2009}.

When the Casimir force between a pair of plates at finite temperature is considered, another source of the force arises - thermal fluctuations of the electromagnetic field. According to Bose-Einstein statistics, the population of an electromagnetic field mode of frequency $\omega$ at temperature $T$ is
\begin{align}
\label{eq:FiniteTemp1}
n(\omega) = \frac{1}{2} + \frac{1}{e^{\hbar\omega/k_B T}-1} = \frac{1}{2}\coth\frac{\hbar\omega}{k_B T},
\end{align}
where the first term describes the zero-point fluctuations, and the second term is the thermal population for a gas of Bose particles (photons); $k_B$ is Boltzmann's constant. This second term gives rise to a finite-temperature term in the Casimir force $F^{(T)}_{\rm C}$. At small separations ($d\lesssim 1$~$\mu$m at $T=300$~K) the thermal force is much smaller than the zero-point force: $F^{(T)}_{\rm C}\ll F^{(T=0)}_{\rm C}$, but for large separations ($d\gtrsim 3$~$\mu$m at $T=300$~K) the thermal force dominates: $F^{(T)}_{\rm C}\gg F^{(T=0)}_{\rm C}$. Although an analogous thermal Casimir-Polder force has been measured between an atom and a surface~\cite{Obrecht2007}, the thermal Casimir force between two macroscopic objects has not been experimentally observed until now.

The magnitude of the force $F^{(T)}_{\rm C}$ has been a subject of intense theoretical debate~\cite{Bostrom2000,Brevik2005,Bezerra2004}. The controversy concerns the form of the behaviour of the plates' complex permittivity $\epsilon(\omega)$ at low frequencies, and the related question of whether or not the zero-frequency transverse-electrical (TE $\omega=0$) mode should be included in the calculation of the finite-temperature force. If the Drude model $\epsilon_{\rm Drude}(\omega)=1-\omega_p^2/\omega(\omega+i\gamma)$ (where $\omega_p$ is the plasma frequency and $\gamma$ is the dissipation rate) is used to describe the low-frequency permittivity, then the TE $\omega=0$ mode does not contribute to the finite-temperature force, whose magnitude at large plate separations is calculated to be (for a sphere-plane geometry with $d\ll R$)
\begin{align}
\label{eq:CasDrude}
F^{(T)}_{\rm C}{\rm (Drude)} = \frac{\zeta(3)}{8}\frac{Rk_BT}{d^2},
\end{align}
where $\zeta$ is the Riemann zeta function, $\zeta(3)\approx 1.202$. The use of the Drude model, however, has been claimed to contradict the third law of thermodynamics (Nernst's heat theorem), as it appears to give rise to finite zero-temperature entropy of the electromagnetic field between the plates, if the plate material dissipation is allowed to vanish in the limit $T \rightarrow 0$, as is the case, for example, in an ideal crystal~\cite{Brevik2005,Bezerra2004} (if the dissipation stays finite as $T \rightarrow 0$, there is no contradiction with the third law). In order to avoid this problem, a different form of the plates' low-frequency permittivity has been proposed: $\epsilon_{\rm plasma}(\omega)=1-\omega_p^2/\omega^2$, this is known as the plasma model. In this model, the TE $\omega=0$ mode does contribute to the finite-temperature force, whose magnitude at large plate separations is double the Drude result:
\begin{align}
\label{eq:CasPlasma}
F^{(T)}_{\rm C} {\rm (plasma)}= \frac{\zeta(3)}{4}\frac{Rk_BT}{d^2}.
\end{align}
Some Casimir force measurements at small plate separations ($d<750$~nm) have been interpreted as being in agreement with the plasma model~\cite{Decca2005}, although at such small separations the relative difference between the Drude and the plasma model predictions is very small.
A calculation of the entropy of the electromagnetic field in the plasma model gives zero at absolute zero temperature,
in agreement with the third law. Let us note, however, that the third law of thermodynamics does allow for a non-vanishing entropy at zero temperature for systems with a degenerate ground state~\cite{Kittel1980}, as is the case with glasses~\cite{Langer1988}. It is possible that the Drude model with vanishing dissipation in the limit $T \rightarrow 0$ leads to a degenerate ground state of the system due to persistent eddy currents (Foucault glass)~\cite{Intravaia2009}.
Equations (\ref{eq:CasDrude}) and (\ref{eq:CasPlasma}) display only the leading temperature-dependent term in the Casimir force; at 300~K and at the larger plate separations considered in our experiment this is the dominant term.

We report the measurement of the finite-temperature contribution to the Casimir force between gold plates.
The total force between the plates measured in our experiment can be written as the sum of the Casimir force $F_C(d)$ and the electrostatic force:
\begin{align}
\label{eq:Electr}
F(d,V) = F_C(d) + \pi \epsilon_0 R \left[ \frac{(V-V_m)^2}{d} + \frac{V_{rms}^2}{d} \right],
\end{align}
where $\epsilon_0$ is the permittivity of free space, and $V$ is the computer-controlled bias voltage applied between the plates. The ``minimizing potential'' offset $V_m$ is due to the contact potential difference of approximately 20~mV between the two plates, caused by the several solder contacts around the electrical loop connecting the two plates.
The second term in brackets is due to the regions (patches) of varying potential on the plate surfaces, caused, for example, by spatial changes in surface crystalline structure, adsorbed impurities, or oxides, and invariably present even on chemically inert metal surfaces prepared in an ultra-clean environment~\cite{Robertson2006,Robertson2007}. The force caused by these patches has been experimentally observed in, for example, Refs.~\cite{Kim2009,Antonini2009}, and is characterized by the parameter $V_{rms}$, related to the magnitude of the voltage fluctuations across the plates.
There are three length scales relevant to the form of the electrostatic patch force: the plate separation $d$, the ``effective interaction length'' $r_{eff}=\sqrt{Rd}$, and the typical patch size $\lambda$.
As shown in Ref.~\cite{Speake2003}, potential patches of size $\lambda\ll d$ lead to an exponentially-suppressed electrostatic force between the plates ($\propto e^{-d/\lambda}$), which we neglect. Potential patches of size $\lambda\gtrsim r_{eff}$ give rise to an electrostatic force of the form $\pi \epsilon_0 R (V_m(d)+V_1)^2/d$, where $V_m(d)$ describes the dependence of the minimizing potential on plate separation $d$, and $V_1$ is a constant~\cite{Kim2010}. We find that for the gold-coated plates used in our experiment $V_m$ is very nearly independent of $d$ (variation is 0.2~mV between 0.7~$\mu$m and 7~$\mu$m), therefore this additional force is small compared to the experimental error, and we do not include it in our data analysis. Finally, potential patches on the length scale $d\ll \lambda \ll r_{eff}$ give rise to an electrostatic force between the plates of the form $V_{rms}^2/d$, which is the last term in Eq.~(\ref{eq:Electr}).

\begin{figure}[h!]
    \includegraphics[width=\columnwidth]{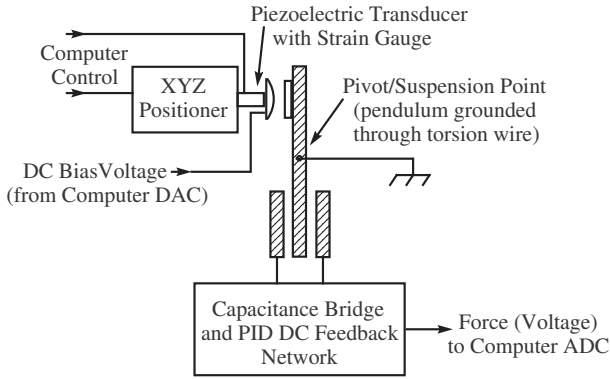}
    \caption{The top-view schematic of the torsion-pendulum experimental apparatus.}
    \label{fig:Setup}
\end{figure}
A top-view schematic of our torsion pendulum apparatus is shown in Figure~\ref{fig:Setup} and described in detail in the Methods section.
The total force between the two plates is measured at 30 logarithmically-spaced plate separations between 0.7~$\mu$m and 7~$\mu$m in a series of 383 sweeps, adding up to a total of 8 days of data taking.
In order to find the plate separation $d$, the force as a function of bias voltage $V$ is recorded at a fixed $d$. A parabolic fit to the force-vs-voltage data is used to extract the separation $d$, and the offset $V_m$. This procedure is repeated at separations of 0.7~$\mu$m and 7~$\mu$m, at intermediate separations the bias voltage is set to $V_m$, eliminating the first term in brackets in equation (\ref{eq:Electr}), and $d$ is determined from the change in the piezoelectric transducer strain gauge reading, pre-calibrated in a separate series of direct measurements. The closest approach of 0.7~$\mu$m was set due to feedback instability at smaller plate separations, caused by the large force gradient~\cite{Lamoreaux1997}.

A systematic correction has to be applied to the data to take into account fluctuations in plate separation $d$~\cite{Lamoreaux2010}. The sources of these fluctuations are surface roughness of the plates, and pendulum fluctuations, caused, for example, by apparatus vibrations. Surface roughness measurements were performed with the Micromap TM-570 interferometric microscope at the Advanced Light Source Optical Metrology Laboratory~\cite{Yashchuk2006, Yashchuk2005}, yielding an rms roughness of $S_q\approx 10$~nm for the curved plate, and $S_q\approx 1$~nm for the flat plate. Vibration-caused fluctuations in $d$ were measured by connecting an inductor in parallel with the Casimir plates, and monitoring the resonance frequency of the resulting LC-circuit; rms fluctuations of $\lesssim 40$~nm were recorded. In addition, statistical error of $\pm 10$~nm in determination of $d$ contributes in quadrature to the fluctuations mentioned above. We take the total rms plate separation fluctuation of $\delta=(40\pm20)$~nm. From the Taylor expansion of the Casimir force about the mean plate separation, we deduce that a correction term $F_C''\delta^2/2$ has to be added to the theoretical force when comparing with experiment, the double prime denotes second-order derivative with respect to $d$. In addition, since the same correction exists for the electrostatic force, the plate separation $d$ extracted from the electrostatic calibration was corrected by a factor $1+(\delta/d)^2$, and the electrostatic patch force $V_{rms}^2/d$ was corrected by the same factor.

\begin{figure}[h!]
    \includegraphics[width=\columnwidth]{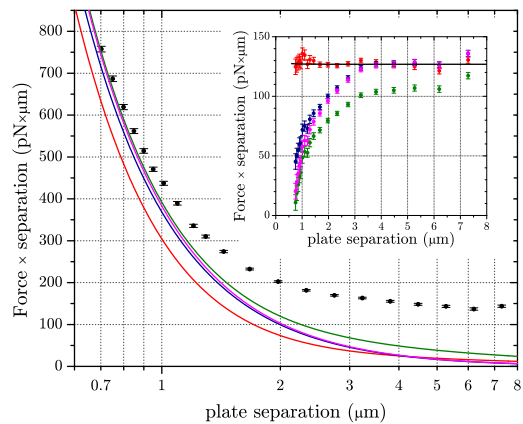}
    \caption{Experimental results for the total short-range force between gold plates. The data have been binned for clarity, the vertical error bars include contributions from the statistical scatter of the points as well as from uncertainties in the applied corrections, discussed in the text. Also shown are the four theoretical models for the Casimir force: (i) the Drude model including the $T=300$~K finite-temperature force (red), (ii) the plasma model including the $T=300$~K finite-temperature force (green), (iii) the Drude model without the finite-temperature force, ie with $T=0$ (blue), and (iv) the plasma model at $T=0$ (magenta). The data are plotted with $F\times d$ on the y-axis, so that the electrostatic force, proportional to $1/d$, appears as a constant offset on the plot. Inset: experimental data with each of the theoretical Casimir force models subtracted, color as above. Electrostatic patch force $\pi \epsilon_0 R V_{rms}^2/d$ corresponds to a constant offset (up to the small $1+(\delta/d)^2$ correction). The fit to the Drude model points is shown by the black line.}
    \label{fig:TotalForce}
\end{figure}
The binned force data, corrected as described above, are shown in Figure~\ref{fig:TotalForce}. These data were taken with the bias potential set equal to the offset $V_m$ (determined at 7~$\mu$m), therefore, as shown in equation (\ref{eq:Electr}), the recorded force is a sum of the Casimir force and the electrostatic patch-potential force, given by the second term in the brackets. Unfortunately an independent measurement of this electrostatic force with the required accuracy is currently not feasible. In order to perform such a measurement, the local surface potential has to be measured with millivolt sensitivity and micron spatial resolution. Commercial Kelvin probes (used for measurements in Ref.~\cite{Robertson2006}, for example) lack the necessary spatial resolution, since, in order to get an appreciable sample-tip capacitance, a tip size of at least 1~mm is needed~\cite{Rossi1992}, and state-of-the-art custom-built Kelvin probes have achieved micron-scale resolution, but have the electrostatic potential sensitivity of only 30~mV~\cite{Cheran2007}.
Therefore the electrostatic patch-potential force has to be modeled and extracted from our data. This is done by fitting the experimental data with the expression of the form $F(d) = F_C(d) + \pi \epsilon_0 R V_{rms}^2/d + a$, where $F_C(d)$ is the theoretical prediction for the Casimir force with no adjustable parameters (see below), and the constant force offset $a$ is due to voltage offsets in measurement electronics; for clarity we subtracted this offset from the displayed data. The two fit parameters $V_{rms}$ and $a$ are the only adjustable parameters used in our data analysis.

We consider four theoretical possibilities for the Casimir force between the gold plates: (i) the Drude model including the $T=300$~K finite-temperature force, (ii) the plasma model including the $T=300$~K finite-temperature force, (iii) the Drude model without the finite-temperature force (ie with $T=0$), and (iv) the plasma model at $T=0$. The Casimir force for each of these models is calculated using the Lifshitz formalism and the gold optical permittivity data~\cite{Palik1998}, extrapolated to zero frequency using the corresponding (Drude/plasma) model with the parameters $\omega_p = 7.54$~eV, $\gamma = 0.051$~eV (details in Methods section). The resulting theoretical force curves are shown in Figure~\ref{fig:TotalForce}. Note that the data are plotted with $F\times d$ on the y-axis, so the patch-potential force contribution leads to a separation-independent offset (up to the small $1+(\delta/d)^2$ correction) of the experimental data points compared to the theoretical models.

\begin{figure}[h!]
    \includegraphics[width=\columnwidth]{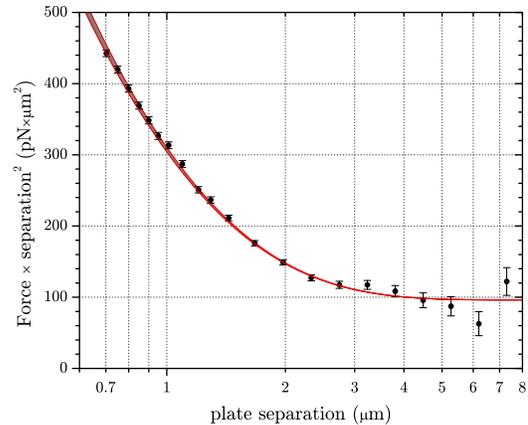}
    \caption{The short-range force data corrected for an electrostatic force with $V_{rms} = 5.4$~mV. The reduced $\chi^2$ of 1.04 demonstrates excellent agreement with the Drude model including the thermal Casimir force at $T=300$~K (red lines). The grey band represents theoretical uncertainty in the Casimir force calculation from the ellipsometry data; the force curve with Drude parameters $\omega_p = 7.54$~eV, $\gamma = 0.051$~eV was chosen for comparison with experiment (see Methods section). The data are plotted with $F\times d^2$ on the y-axis, so that the thermal Casimir force, given by equation (\ref{eq:CasDrude}), corresponds to an offset of 97~pN$\times\mu$m$^2$, which dominates the force at large plate separations. In this region the Casimir force is largely independent of the material properties of the plates.}
    \label{fig:Drude}
\end{figure}
The fit $F(d) = F_C(d) + \pi \epsilon_0 R V_{rms}^2/d + a$ is performed independently for the four theoretical models. The best agreement is obtained for the Drude model at $T=300$~K, with the reduced $\chi^2$ of 1.04. The rms patch potential fluctuation obtained from the fit is $V_{rms} = (5.4\pm0.1)$~mV, consistent with the magnitude of potential fluctuations expected across a gold surface due, for example, to work function variations~\cite{Speake2003}. The force offset obtained from the fit is $a=(-3.0\pm 0.4)$~pN. Having extracted the electrostatic contribution to the force, and the offset, we can subtract them from the experimental data, the remainder is the Casimir force, plotted in Figure~\ref{fig:Drude} together with the theoretical prediction of the Drude model at $T=300$~K. We plot $F\times d^2$ on the y-axis, so that the $1/d^2$ finite-temperature force of equation (\ref{eq:CasDrude}) corresponds to an offset that dominates the force at large plate separations ($d\gtrsim 3$~$\mu$m). The experiment is in excellent agreement with the Drude model containing the thermal Casimir force. Note that the theoretical calculation of the thermal Casimir force, that dominates at large separations, is largely independent of the exact values of the Drude parameters.

The experimental data rules out the other three theoretical Casimir force models. The reduced $\chi^2$ obtained from the fit is: (ii) $\chi^2=32$ for the plasma model at $T=300$~K (best fit $V_{rms} = 3.0$~mV), (iii) $\chi^2=23$ for the Drude model without the $T=300$~K temperature correction (best fit $V_{rms} = 4.0$~mV), and (iv) $\chi^2=43$ for the plasma model without the $T=300$~K temperature correction (best fit $V_{rms} = 3.6$~mV). Therefore our experiment rules out the plasma model for the Casimir force between gold plates in the separation range 0.7~$\mu$m to 7~$\mu$m, confirms the Drude model, and demonstrates the existence of the $T=300$~K thermal Casimir force. The thermal Casimir force drops off as $T/d^2$, and therefore dominates over the purely quantum $T=0$ Casimir force (which behaves roughly as $1/d^3$) for plate separations greater than 3~$\mu$m.


\section{Methods}

A torsion pendulum is suspended inside a vacuum chamber (pressure $5\times10^{-7}$~torr) by a tungsten wire of 25~$\mu$m diameter and 2.5~cm length. The force to be measured is between the two ``Casimir plates'', each coated with a 700~\AA~(optically thick) layer of gold evaporated on top of a 100~\AA-thick layer of titanium. One is a flat plate mounted on one side of the pendulum, as shown in Figure~\ref{fig:Setup}, the other is a spherical lens (radius of curvature 15.6~cm, measured with the Micromap TM-570 interferometric microscope at the Advanced Light Source Optical Metrology Laboratory together with surface roughness, and found to vary by less than 2\% over the surface of the lens), mounted on a Thorlabs T25 XYZ positioning stage, which, together with a piezoelectric transducer, is used to vary the plate separation $d$. The attractive force between the plates creates a torque on the pendulum body, which is counteracted by a pair of ``compensator'' electrodes on the opposite end of the pendulum. The voltage that has to be applied to the compensator electrodes to keep the pendulum stationary is proportional to the force between the Casimir plates, with the calibration coefficient extracted from the measurements of the electrostatic force between the plates. Pendulum rotation is detected by a capacitance bridge connected to the compensator electrodes, and a servo loop is used to apply the compensator electrode voltage necessary to hold the pendulum in equilibrium (at a fixed angle). A NdFeB magnet is placed under the pendulum body to damp the swinging modes of the pendulum. The experiment is placed on a vibration-isolation slab, extending down to the bedrock below the building foundation. Subsequent measurements indicated that a smooth long-term drift, which is subtracted from the experimental data, is due to tilting of this slab
(measured by an Applied Geomechanics model 701-2A tilt meter), correlated with ambient humidity. Both humidity and temperature were monitored using an Onset HOBO U12 data logger. Temperature variations were less than 1$^{\deg}$C.

Ellipsometric measurements in the wavelength range 191~nm to 1700~nm were performed on gold-coated plates prepared identically to the ones used in the force measurement apparatus. The resulting absorption and reflectivity data were within a few percent of the data given in Ref.~\cite{Palik1998}. In order to calculate the theoretical Casimir force with the Lifshitz formalism, the data were extrapolated to higher frequencies using the data in Ref.~\cite{Palik1998}, and to zero frequency using either the Drude or the plasma model. The following ranges for the plasma frequency and dissipation were explored: $\omega_p = 6.85$~eV to 9.00~eV, and $\gamma = 0.02$~eV to 0.061~eV,
resulting in theoretical force variation of up to 3\%, shown by the grey band in Figure~\ref{fig:Drude}. The parameters $\omega_p = 7.54$~eV, $\gamma = 0.051$~eV were chosen for the theoretical Casimir force that was compared with experimental data, since they best fit the optical data given in Ref.~\cite{Palik1998}.

\section*{Acknowledgements}
The authors thank Valery Yashchuk for performing the surface roughness measurements, and acknowledge discussions with Stephen Eckel and  Francesco Intravaia. This work was supported by the DARPA/MTOs Casimir Effect Enhancement project under SPAWAR Contract No. N66001-09-1-2071.

\bibliography{References}

\end{document}